\documentclass[10pt,letterpaper,twocolumn]{article} 

\usepackage{ol2}
\usepackage[draft]{hyperref}
\usepackage{amsmath}
\usepackage{color}

\begin{document}

\twocolumn[ 

\title{
Decoy-State Quantum Key Distribution with Nonclassical Light \\Generated in a One-dimensional Waveguide
}
\author{Huaixiu Zheng$^{*}$, Daniel J. Gauthier, and Harold U. Baranger$^{\dagger}$}
\address{Department of Physics, Duke University, P. O. Box 90305,
Durham, North Carolina 27708, USA\\
$^*$Corresponding author: hz33@duke.edu\\
$^{\dagger}$Corresponding author: baranger@phy.duke.edu
}


\begin{abstract}
We investigate a decoy-state quantum key distribution (QKD) scheme with a sub-Poissonian single-photon source, which is generated on demand by scattering a coherent state off a two-level system in a one-dimensional waveguide. 
We show that, compared to coherent state decoy-state QKD, there is a two-fold increase of the key generation rate. 
Furthermore, the performance is shown to be robust against both parameter variations and loss effects of the system.
\end{abstract}


 ] 

\noindent Quantum key distribution (QKD), which allows two distant users, Alice and Bob, to share a secret key with security guaranteed by principles of quantum physics,
is the first commercially available application of quantum information science.
The first QKD protocol, BB84 \cite{BB84}, proposes to use an ideal single-photon source, which is still beyond the current technology despite tremendous experimental effort worldwide.
Hence, most QKD experiments use weak coherent states (WCS) from attenuated lasers as a photon source \cite{ScaraniRMP09, LoPRL12}.
Two drawbacks come with the WCS: the multiphoton and the vacuum components. 
The vacuum content limits Bob's detection rate, and hence leads to a shorter maximal distance.
The multiphoton component makes QKD vulnerable to the photon number splitting attack, where the eavesdropper (Eve) can suppress single-photon signals and split multiphoton signals, keeping one copy and sending one copy to Bob.
This way, Eve obtains the full information without being detected, and the unconditional security breaks down.
The decoy state method was proposed to beat such attacks \cite{HwangPRL03, LoPRL05, WangPRL05}: 
Alice prepares additional decoy states, and learns about the eavesdropping from their transmission.
Recently, alternative light sources, including spontaneous parametric down-conversion \cite{AdachiPRL07} and heralded single-photons \cite{WangPRL08}, have been used in decoy-state QKD.

In this paper, we combine the decoy-state method with a sub-Poissonian single-photon source generated on demand by scattering in a waveguide. We find that there is a substantial increase in the key generation rate and maximal transmission distance compared to both WCS and heralded single-photon decoy-state QKD. Furthermore, the performance is robust against either parameter variation or loss in the system, making it a promising candidate for future QKD systems.


Recently, strong coupling between light and matter has been achieved in a variety of one-dimensional (1D) waveguide-QED systems \cite{ClaudonNatPhoton10, AstafievSci10,HoiPRL12,LauchtPRX12}.
This great experimental progress has stimulated extensive theoretical study of nonlinear effects in such systems for the purpose of quantum information processing \cite{ChangNatPhy07,ShenPRL07,RephaeliPRA11,RoyPRL11,KolchinPRL11,ZhengPRL11}.
One example, on which we base this work, is to generate nonclassical light \cite{ZhengPRA10,HoiPRL12} by sending a coherent state into a 1D waveguide which is side-coupled to a quantum nonlinear element, such as a two-level system (2LS) \cite{ChangNatPhy07,ShenPRL07}.
The nonlinearity of the quantum element leads to a distinct difference between multiphoton and single-photon scattering.
For example, when two photons interact with a 2LS simultaneously, the 2LS will only be able to absorb one photon and hence the pair will have a high transmission probability,
leading to photon bunching and super-Poissonian photon statistics \cite{ZhengPRA10,HoiPRL12}. 
Here, we focus on the reflected field, and show that it has sub-Poissonian statistics. 

\begin{figure}[tb]
\centering
 \includegraphics[width=0.41\textwidth]{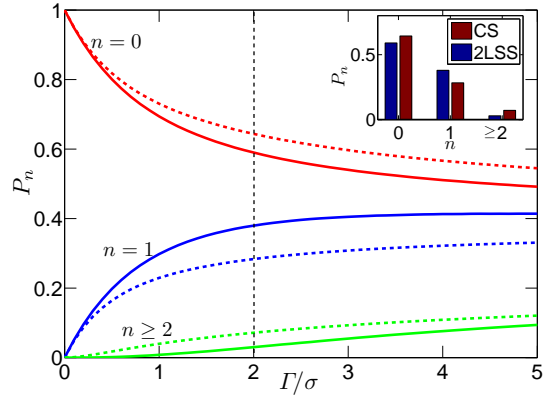}
\caption{(Color online) Nonclassical light source. 
The number statistics $P_n$ of the 2LS source (2LSS, solid), and a coherent state (CS, dashed) of the same mean photon number as a function of $\Gamma/\sigma$.
Inset: $P_n$ at $\sigma=\Gamma/2$. Here, we set $P=\Gamma/\Gamma^{\prime}=20$.}
\label{Figure1}
\end{figure}

Figure\,\ref{Figure1} shows the probabilities $P_n$ 
to measure $n$-photon states in the reflected field after scattering a coherent state off the 2LS.
We will call such a photon source the ``2LS source''.
The input coherent state has mean photon number $\bar{n}=1$. 
We take a Gaussian wave-packet with central frequency on resonance with the 2LS and root-mean-square spectral width $\sigma$.
The model we use is presented in detail in Ref.\,\onlinecite{ZhengPRA10}. It includes absorption of photons by the 2LS and subsequent spontaneous emission into the waveguide mode at rate $\Gamma$.
In addition, 
a loss rate $\Gamma^{\prime}$ models subsequent emission into modes other than the waveguide mode (as well as possible non-radiative processes) \cite{ChangNatPhy07,ZhengPRA10}.
The loss rate $\Gamma^{\prime}$ decreases the number of photons in the waveguide, thus changing the number statistics. 
When $\Gamma^{\prime}$ is small compared to $\sigma$ and $\Gamma$, multi-photon loss is negligible: the loss from an $n$-photon pulse appears as an $n-1$ photon pulse, 
for which we then naturally use the $n-1$ photon transmission and reflection to distribute the probability between the transmitted and reflected fields.

We set the effective Purcell factor $P=\Gamma/\Gamma^{\prime}=20$ \cite{LauchtPRX12} for now, and return to the effect of loss later.
For comparison, we also show $P_n$ (dashed line) of a coherent state with the same mean photon number as the reflected field. 
It is remarkable that, for the full parameter range, the reflected field has  higher single-photon and lower vacuum and multiphoton content than the coherent state.
In the insert of Fig.\,\ref{Figure1}, we show that the multiphoton content is strongly suppressed at $\sigma=\Gamma/2$. 
This is in agreement with the observed antibunching behavior of microwave photons \cite{HoiPRL12}, and is the key to increasing the key generation rate.

Now, we discuss the decoy-state method with light sources, including weak coherent states, a heralded single-photon source (HSPS), and the 2LS source (2LSS).
The secure key generation rate (per signal pulse emitted by Alice) is given by \cite{GottesmanQIC04}
\begin{equation}
 R\geq q\{-Q_s f(E_s)H_2(E_s)+Q_1[1-H_2(e_1)]\},
\label{Eq1}
\end{equation}
where the efficiency $q$ is $1/2$ for the Bennett-Brassard 1984 (BB84) protocol, 
$f(E_s)$ is the error correction efficiency (we use $f=1.22$ \cite{BrassardPRL00}),
$Q_s$ and $E_s$ are the overall gain and error rate of signal states, respectively,
$Q_1$ and $e_1$ are the gain and error rate of single-photon states, respectively,
and $H_2(x)$ is the binary Shannon information function: $H_2(x)=-x\text{log}_2(x)-(1-x)\text{log}_2(1-x)$. 

In Eq.\,(\ref{Eq1}), while $Q_s$ and $E_s$ are measurable quantities in experiments, $Q_1$ and $e_1$ are unknown variables.
$Q_s$ and $E_s$ are given by
\begin{equation}
 Q_s=\sum_{n=0}^{\infty}p_n^sY_n,\quad
 E_s=\frac{1}{Q_s}\sum_{n=0}^{\infty}p_n^sY_ne_n,
\label{Eq2}
\end{equation}
where $p_n^s$ is the $n$-photon probability of signal states, 
$e_n$ is the error rate of an $n$-photon state,
and $Y_n$ is the $n$-photon yield, i.e., the conditional probability of a click on Bob's side given that Alice has sent an $n$-photon state.

To generate a lower bound on the key generation rate, we have to estimate a lower bound of $Q_1$ (or equivalently $Y_1$ as $Q_1=p_1^sY_1$) and an upper bound of $e_1$.
Estimating the lower bound $Y_1^l$ and the upper bound $e_1^u$ based solely on Eq.\,(\ref{Eq2}) unavoidably underestimates the secure key generation rate due to the lack of enough information about the transmission channel.
The decoy-state idea \cite{HwangPRL03, LoPRL05, WangPRL05} is a clever way to obtain additional channel information by sending in additional decoy states.
The decoy states are used to detect eavesdropping, but not for key generation. By measuring the transmission of the decoy states, Alice and Bob have another set of constraints
\begin{equation}
 Q_d=\sum_{n=0}^{\infty}p_n^dY_n,\quad
 E_d=\frac{1}{Q_d}\sum_{n=0}^{\infty}p_n^dY_ne_n,
\label{Eq3}
\end{equation}
where $Q_d$ and $E_d$ are the measured overall gain and error rate of decoy states, respectively. 
Because Eve has no way to distinguish an $n$-photon decoy state from an $n$-photon signal state, the yield $Y_n$ and the error rate $e_n$ are the same for both the decoy and signal states. 

For our numerical simulation, we use the channel model in Ref.\,\onlinecite{MaPRA05} to calculate the experimental parameters $Q_s$, $E_s$, $Q_d$, and $E_d$.
In this model, the yield is $Y_n=1-(1-Y_0)(1-\eta)^n$, where $Y_0$ is the background rate and $\eta$ is the overall transmittance given by 
$\eta=t_{AB}\eta_{Bob}$, where $t_{AB}=10^{-\alpha \ell/10}$ is the channel transmittance and $\eta_{Bob}$ is the detection efficiency on Bob's side. 
Here, $\alpha$ is the loss coefficient and $\ell$ is the transmission distance.
The error rate is given by $e_n=[e_0Y_0+e_d(Y_n-Y_0)]/Y_n$, where $e_d$ is the probability that a photon hits the wrong detector and $e_0$ is the error rate of the background.
We use the experimental parameters in Ref.\,\onlinecite{GobbyAPL04}: $\alpha=0.21\text{dB/km}$, $e_d=3.3\%$, $Y_0=1.7\times10^{-6}$, $e_0=0.5$, and $\eta_{Bob}=0.045$.

\begin{figure}[b]
 \centering
 \includegraphics[width=0.41\textwidth]{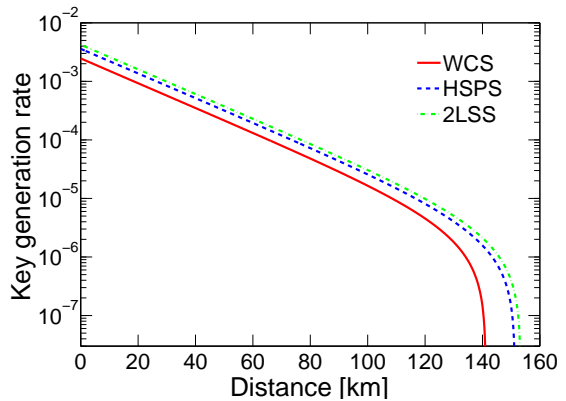}
 \caption{(Color online) Key generation rate with different light sources: weak coherent state (WCS), heralded single-photon source (HSPS), and 2LS source (2LSS) with $\sigma=\Gamma/2$ and $P=20$.}
 \label{Figure2}
\end{figure}

\begin{figure}[tb!]
 \centering
 \includegraphics[width=0.38\textwidth]{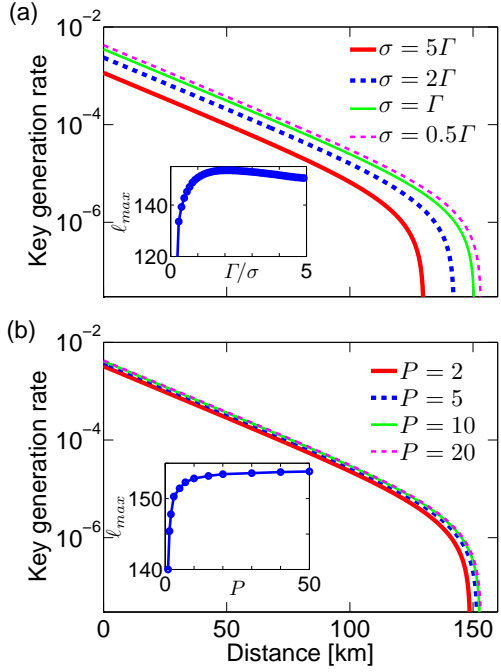}
 \caption{(Color online) Key generation rate of 2LS sources with parameter variation and loss: 
                            (a) $\sigma=5\Gamma, 2\Gamma, \Gamma, 0.5\Gamma$. Inset: maximal transmission distance ($\ell_{\text{max}}$) as a function $\Gamma/\sigma$. $P=\Gamma/\Gamma^{\prime}=20$;
			    (b) $P=2, 5, 10, 20$. Inset: $\ell_{\text{max}}$ as a function of $P$. $\sigma=\Gamma/2$.}
 \label{Figure3}
\end{figure}

We apply the linear programming method \cite{CLRS09} to estimate $Y_1^l$ and $e_1^u$ from Eqs.\,(\ref{Eq2}) and (\ref{Eq3}). 
This method is applicable to light sources with general number statistics. 
We use two decoy states---the vacuum and a weak decoy state.
For the weak coherent states, the key generation rate is optimized in terms of the mean photon number in both the signal and decoy states \cite{MaPRA05}.
For the heralded single-photons, we use the number statistics from Ref.\,\onlinecite{WangPRL08}.
For the 2LS source, the signal and decoy states are generated by scattering coherent states of $\bar{n}=1$ and $\bar{n}=0.02$, respectively. 
We choose $\sigma=\Gamma/2$.

Figure\,\ref{Figure2} shows the resulting key generation rate.
With the same experimental parameters and estimation technique, our scheme using the 2LS source obtains a two-fold increasing of key generation rate compared to the WCS method.
The maximal transmission distance is increased as well. In addition, our scheme also outperforms the HSPS scheme. 
Such a performance enhancement is due to the reduced vacuum and multiphoton contents, as shown in Fig.\,\ref{Figure1}.

Next, we investigate the robustness of our scheme with respect to the variation of system parameter $\Gamma/\sigma$. 
As shown in Fig.\,\ref{Figure3}(a), the key generation rate gradually converges as $\Gamma/\sigma$ increases. 
In particular, the insert shows that the maximal transmission distance ($\ell_{\text{max}}$) has little change for $\Gamma/\sigma\geq1$.

The effect of loss on the system performance is shown in Figure\,\ref{Figure3}(b). We fix $\sigma=\Gamma/2$ and choose different loss rates $\Gamma^{\prime}$ of the 2LS. 
In Fig.\,\ref{Figure3}(b), we observe that, as $P$ increases, the key generation rate increases and converges.
It is evident that, for $P\geq10$, the performance is very reliable against loss as shown in the insert. 
Given that values of $P$ as large as $20$ have already been achieved in recent experiments \cite{LauchtPRX12,MikkelsenPrivate12}, our scheme can be practically useful for quantum key distribution.


This work was
supported by US NSF\,Grant\,No.\,PHY-10-68698. H.Z.\ is
supported by a John T.\ Chambers Fellowship from the Fitzpatrick
Institute for Photonics at Duke University.


\begin{thebibliography}{10}
\newcommand{\enquote}[1]{``#1''}

\bibitem{BB84}
C.~H. Bennett and G.~Brassard, \enquote{Proceedings of Ieee International
  Conference on Computers, Systems, and Signal Processing,}  (IEEE, New York,
  1984), pp. 175--179.

\bibitem{ScaraniRMP09}
V.~Scarani, H.~Bechmann-Pasquinucci, N. J.~Cerf, M.~Du\ifmmode \check{s}\else \v{s}\fi{}ek, N.~L\"utkenhaus, and M.~Peev,
\enquote{The security of practical quantum key distribution,} Rev. Mod. Phys. \textbf{81}, 1301 (2009).


\bibitem{LoPRL12}
H.-K.~Lo, M.~Curty, and B.~Qi, \enquote{Measurement-Device-Independent Quantum Key Distribution,} Phys. Rev. Lett. \textbf{108}, 130503 (2012).

\bibitem{HwangPRL03}
W.-Y. Hwang, \enquote{Quantum key distribution with high loss: Toward global
  secure communication,} Phys. Rev. Lett. \textbf{91}, 057901 (2003).


\bibitem{LoPRL05}
H.-K.~Lo, X.~Ma, and K.~Chen,\enquote{Decoy State Quantum Key Distribution,} Phys. Rev. Lett. \textbf{94}, 230504 (2005).

\bibitem{WangPRL05}
X.-B.~Wang, \enquote{Beating the Photon-Number-Splitting Attack in Practical Quantum Cryptography,} Phys. Rev. Lett. \textbf{94}, 230503 (2005).

\bibitem{AdachiPRL07}
Y.~Adachi, T.~Yamamoto, M.~Koashi, and N.~Imoto, \enquote{Simple and efficient
  quantum key distribution with parametric down-conversion,} Phys. Rev. Lett.
  \textbf{99}, 180503 (2007).

\bibitem{WangPRL08}
Q.~Wang, W.~Chen, G.~Xavier, M.~Swillo, T.~Zhang, S.~Sauge, M.~Tengner, Z.-F.
  Han, G.-C. Guo, and A.~Karlsson, \enquote{Experimental decoy-state quantum
  key distribution with a sub-poissionian heralded single-photon source,} Phys.
  Rev. Lett. \textbf{100}, 090501 (2008).

\bibitem{ClaudonNatPhoton10}
J.~Claudon, J.~Bleuse, N.~S. Malik, M.~Bazin, P.~Jaffrennou, N.~Gregersen,
  C.~Sauvan, P.~Lalanne, and J.-M. G\'{e}rard, \enquote{A highly efficient
  single-photon source based on a quantum dot in a photonic nanowire,} Nat.
  Photon. \textbf{4}, 174 (2010).

\bibitem{AstafievSci10}
O.~Astafiev, A.~M. Zagoskin, A.~A. Abdumalikov, Y.~A. Pashkin, T.~Yamamoto,
  K.~Inomata, Y.~Nakamura, and J.~S. Tsai, \enquote{{Resonance Fluorescence of
  a Single Artificial Atom},} Science \textbf{327}, 840--843 (2010).

\bibitem{HoiPRL12}
I.-C. Hoi, T.~Palomaki, J.~Lindkvist, G.~Johansson, P.~Delsing, and C.~M. Wilson
\enquote{Generation of Nonclassical Microwave States Using an Artificial Atom in 1D Open Space,} Phys. Rev. Lett. \textbf{108}, 263601 (2012).


\bibitem{LauchtPRX12}
A.~Laucht, S.~P\"utz, T.~G\"unthner, N.~Hauke, R.~Saive, S.~Fr\'ed\'erick,
  M.~Bichler, M.-C. Amann, A.~W. Holleitner, M.~Kaniber, and J.~J. Finley,
  \enquote{A waveguide-coupled on-chip single-photon source,} Phys. Rev. X
  \textbf{2}, 011014 (2012).

\bibitem{ChangNatPhy07}
D.~E. Chang, A.~S. S\o{}rensen, E.~A. Demler, and M.~D. Lukin, \enquote{A
  single-photon transistor using nanoscale surface plasmons,} Nature Phys.
  \textbf{3}, 807--812 (2007).

\bibitem{ShenPRL07}
J.-T. Shen and S.~Fan, \enquote{Strongly correlated two-photon transport in a
  one-dimensional waveguide coupled to a two-level system,} Phys. Rev. Lett.
  \textbf{98}, 153003 (2007).

\bibitem{RephaeliPRA11}
E.~Rephaeli, S.~E. Kocabas, and S.~Fan, \enquote{Few-photon transport in a
  waveguide coupled to a pair of colocated two-level atoms,} Phys. Rev. A
  \textbf{84}, 063832 (2011).

\bibitem{RoyPRL11}
D.~Roy, \enquote{Two-photon scattering by a driven three-level emitter in a
  one-dimensional waveguide and electromagnetically induced transparency,}
  Phys. Rev. Lett. \textbf{106}, 053601 (2011).

\bibitem{KolchinPRL11}
P.~Kolchin, R.~F. Oulton, and X.~Zhang, \enquote{Nonlinear quantum optics in a
  waveguide: Distinct single photons strongly interacting at the single atom
  level,} Phys. Rev. Lett. \textbf{106}, 113601 (2011).

\bibitem{ZhengPRL11}
H.~Zheng, D.~J. Gauthier, and H.~U. Baranger, \enquote{Cavity-free photon
  blockade induced by many-body bound states,} Phys. Rev. Lett. \textbf{107},
  223601 (2011).

\bibitem{ZhengPRA10}
H.~Zheng, D.~J. Gauthier, and H.~U. Baranger, \enquote{Waveguide qed: Many-body
  bound-state effects in coherent and fock-state scattering from a two-level
  system,} Phys. Rev. A \textbf{82}, 063816 (2010).

\bibitem{GottesmanQIC04}
D.~Gottesman, H.-K. Lo, N.~Lutkenhaus, and J.~Preskill, \enquote{Security of
  quantum key distribution with imperfect devices,} Quantum Inf. Comput.
  \textbf{4}, 325 (2004).

\bibitem{BrassardPRL00}
G.~Brassard, N.~L\"utkenhaus, T.~Mor, and B.~C. Sanders, \enquote{Limitations
  on practical quantum cryptography,} Phys. Rev. Lett. \textbf{85}, 1330--1333
  (2000).

\bibitem{MaPRA05}
X.~Ma, B.~Qi, Y.~Zhao, and H.-K. Lo, \enquote{Practical decoy state for quantum
  key distribution,} Phys. Rev. A \textbf{72}, 012326 (2005).

\bibitem{GobbyAPL04}
G.~Gobby, Z.~L. Yuan, and A.~J. Shields, \enquote{Quantum key distribution over
  122 km of standard telecom fiber,} Appl. Phys. Lett. \textbf{84}, 3762--3764
  (2004).

\bibitem{CLRS09}
T.~H. Cormen, C.~E. Leiserson, R.~L. Rivest, and C.~Stein, \emph{Introduction
  to Algorithms} (MIT Press and McGraw-Hill, New York, 2009), 3rd ed.

\bibitem{MikkelsenPrivate12}
M.~H. Mikkelsen, private communication (2012). 

\end{thebibliography}

%
%

\pagebreak

\end{document}